\documentclass[12pt]{article}
\usepackage{epsfig}
\usepackage{amsfonts}
\usepackage{amsopn}
\usepackage{amsmath}
\usepackage{verbatim,amsthm}
\title
{\vskip -50 pt
\begin{flushright}
\normalsize\rm NORDITA 2022-027
\end{flushright}
\vskip 20 pt
On possible composite structure of scalar fields in expanding universe 
}

\author{
 A. A. Zheltukhin 
 \thanks{E-mail: aaz@nordita.org}  \\
Kharkov Institute of Physics and Technology, \\
1, Akademicheskaya St., Kharkov, 61108, Ukraine \\  
NORDITA, KTH Royal Institute of Technology and  \\Stockholm University,
Hannes Alfv\'{e}ns v\"{a}g 12, \\
SE-106 91 Stockholm, Sweden 
}

\date{}
\begin{document}
\maketitle

\begin{abstract}

 Scalar fields in curved backgrounds are assumed to be composite objects. As an example realizing such a possibility we consider a model of the massless tensor field $l_{\mu\nu}(x)$ in a 4-dim. background $g_{\mu\nu}(x)$ with spontaneously broken Weyl and scale symmetries. It is shown that the potential of $l_{\mu\nu}$, represented by a scalar quartic polynomial, has the degenerate extremal described by the composite Nambu-Goldstone scalar boson $\phi(x):=g^{\mu\nu}l_{\mu\nu}$.
Removal of the degeneracy shows that $\phi$ acquires a non-zero vev $\langle\phi\rangle_{0}=\mu$ which, together with the free parameters of the potential, defines the cosmological constant. The latter is zero for a certain choice of the parameters. 
\end{abstract}
\bigskip 

Our universe expands and this process has two certain periods.
 The inflation period was very short and lasted from $10^{-36}$ to $10^{-32}$ 
 seconds after the Big Bang \cite{Guth, Strb, Lind, Stein}. 
 The period of accelerated expansion began eight billion years later and is continuing on now. Both are conjectered as caused by hypothetic quantum fields with special properties of their vacua.  
Inflation posits inflaton - a scalar field with the specific shape of its potential which provides a non-zero vacuum expectation value for inflaton. At the end of the slow-rolling regime inflaton reaches the bottom of its potential. This evolution picture matches the exponentially expanding de Sitter universe and the Standard Model \cite{DaR}, (\cite {Lids} and refs. there). The assumed slow-rolling constraint requires fine-tuning of the initial data with the postulates of quantum field theory and general relativity. Although the iflation predicts and explains many observational data it is not known yet what powered it. 
 
The period of accelerated expansion is explained by the presence of dark energy which is homogeneously distributed across all space-time. In the period of early universe the effect of dark energy was negligibly small but its effect becomes dominant at very long distances 
\cite{Peebl, CDS, Planck, LKRS}.
 Dark energy creates a repulsive force which dominates the attractive gravitational force of visible and dark matter responsible for decelerated expansion of the universe. It is believed that dark energy is bound up with the cosmological constant encoding vacuum energy of the space-time 
 \cite{Carr, Kra}. A hard problem here is a huge discrepancy between the experimental value of the cosmological constant and vacuum energy predicted by quantum field theory \cite{StTu, Dolg, BD}. 
 It was proposed to treat dark energy as a vacuum energy called quintessence that originated  from potential energy of very light dynamical particles \cite{Wett, ChPo, CSSX}.
 They are associated with a scalar field creating the fifth repulsive force that causes the accelerated expansion \cite{DvZ, Bol, LKRDS}. The latter is generally predicted to be slightly less than the one given by the cosmological constant. A special case is phantom energy which is the quintessence energy characterized by the density increasing  in time. Such a case may cause the expansion that exceeds the speed of light resulting in improbable scenario in which all interactions vanish through a finite time \cite{Cald}. 
Modified gravity models suppose alternative way which does not need to use the dark energy image \cite{DE, Zwi, LPPS, SMy, LoLi}. 
 It is amazing that for each of the complicated evolutional periods on different energy scales, 
 the use of scalar fields such as inflaton or quintessence boson has been successful in getting 
 closer to understanding of the dynamics of expanded universe. 
 A key element in this picture seems to be connected with the conception of spontaneously broken symmetries caused by degeneration of vacua that trigger creation of the Nambu-Goldstone (N-G) bosons. 
Scalar bosons were also predicted by the Standard Model and string theory, but the Higgs boson was the only fundamental scalar field observed at the current energies of the LHC. This boson is responsible for the electroweak symmetry breaking. So, it was natural to identify the Higgs boson with the inflaton \cite{BeSh, DeHeWi}. However, it soon became clear that this assumption violated the perturbative unitarity condition \cite{LeMD, BuLeT}. 
Attempts to overcome this problem  showed \cite{AtCa} that the Higgs boson alone was not able to yield the dynamics mediated by the inflaton without adding new degrees of freedom \cite{GL}. 
Such  scalar DOF arise in the supergravity model unifying the Volkov-Akulov \cite{VA} and a chiral superfield \cite{ADFS}. 
 A wider set of new DOF desribed by the massless chiral $\mathcal{N}=1 \,  D=4$ and $\mathcal {N}=1 \,  D=10$ multiplets $(s,s+1/2)$ with spins  $(s=0, 1/2, 1, 3/2, 2,...)$ emerges from the $\theta$-twistor description \cite{Ztw, Ber, CTZ}.
Spontaneous breaking of the local supersymmetry points to degeneration among the extremals of the effective quartic Higgs potential in supergravity models \cite{FNNT}. A hierarchy between the Planck mass, cosmological constant and electroweak scale was explained by this degeneration. 
Detection of new DOF in accelerators and cosmic experiments on TEV scales may lead to new physics outside the SM.  Scalar fields which arise in phenomenological models of systems with spontaneously broken symmetries  identified with the (pseudo)N-G bosons \cite{PQ, Kim, DiFiS, WiGGl, DaWa, Wil, FFO}. 
Axion and  $\pi$-mesons give well-known examples of such bosons arising due to spontaneous breaking of the approximate $U(1)_{axial}$ and chiral symmetries, where $\pi$-mesons are considered as composed by  $q\bar{q}$ pairs.
This hints that scalar fields in cosmology may also have a composite structure formed by the metric  field $g_{\mu\nu}(x)$ of a curved background and hypothetic tensor fields. 
The latter may originate from quark-gluon plasma and have their own internal multiquark structure desribed a global symmetry $G$. 
Study of tensor models invariant under the group $G$ and diffeomorphisms may help, in particular, 
to understand the nature of the inflaton and dark matter. 
 On this way it's interesting to find such nonlinear interaction potentials of the tensors that would have nontrivial extremals. The latter together with the equations of motion (EOM) could indicate on the presence of spontaneous breaking of the symmetry group $G$.
We suppose that the extremals of the potential are expressed in terms of covariant effective fields composed of the metric and tensor fields. 
Then the corresponding  condensates of the tensor fields will encode effective modes of excitations of quark-gluon plasma. Cooling of plasma rebuilds the tensors of the previous stage 
due to changes of thermodynamic scales. This deforms the interaction potential and its extremals defining non-zero vev of the composite fields. 
If these fields appear in the form of diffeomorphism-invariant combinations of $g_{\mu\nu}$ and tensors, then they can be seen as composite scalars.
An example which sustains the above scenario in a $4$-dim. background is proposed here.
For this purpose we use the known idea to treat the global scale symmetry of our universe as spontaneously broken symmetry \cite{Adl, Smol, Zee, BST, RV}. It is in agreement with the experimental data of Planck \cite{Planck} and other high precision experiments. 

In previous known models the scale symmetry breaking is achieved by using a scalar field $\phi$ of the Brans-Dicke type that is understood to represent an elementary particle with zero spin. The corresponding potential $V \propto \lambda\phi^{4}$ which is added in a gravity action quadratic in curvature. 
Here we study an action invariant under the global Weyl and scale symmetries that  
includes a symmetric massless tensor $l_{\mu\nu}$ instead of the scalar $\phi$. 
In the particle physics such a tensor is understood as an elementary particle carrying any spin from the set s=2, 1, 0. The proposed model reveals that $l_{\mu\nu}$ encodes a composite scalar field carrying spin zero. It is notable that this result holds for the general polynomial potential admissible with the invariance of the action under diffeomorphisms and scale transformations. 

Therefore, the presence in the proposed model of two massless symmetric tensors $l_{\mu\nu}$ and $g_{\mu\nu}$ does not break the well known no-go theorem in the theory of gravity that forbids the presence of multiple massless particles with spins equal to 2. 
A hint of this result follows also from the existence of a natural geometric scalar $\phi= g^{\mu\nu}l_{\mu\nu}$, where the tensor field encodes the degrees of freedom of an evolutional period of the expanding universe.
The potential of $l_{\mu\nu}$ consistent with the symmetries of the studied action 
is given by the general homogenious quartic polynomial, containing a few dimensionless parameters, 
that generalizes the $\lambda\phi^{4}$ potential, 
Study of critical points of the potential of $l_{\mu\nu}$ reveals its non-trivial extremal $l_{0\mu\nu}(x)= (\phi_{0}(x)/4)g_{\mu\nu}$, where $\phi_{0}=l_{0\mu\nu}g^{\mu\nu}$. 
This extremal preserves all symmetries of the action and forms an infinitely degenerate vacuum manifold 
\footnote{Assumption that the parameters of spontaneously symmetry breaking can be higher rank tensors was previously pointed out in \cite{VZB}.}. 
Substitution of $l_{0\mu\nu}$ into the EOM for $l_{\mu\nu}$ shows the presence of the particular solution $\phi_{o}=\mu=constant$. This solution removes infinite degeneration of the vacuum manifold 
described by the order parameter $\phi_{0}(x)$ forming a linear representation of the Weyl and scale 
 groups. The removal of this degeneration is realized by choosing a fixed value 
 of the function $\phi_{0}(x)$ equal to $\phi_{o}$. This choice is interpreted as fixing the vacuum state.  Pictorially this fixing is represented by using two different subscripts "0" and "o".
Thereby it is realized spontaneous breakdown of the scale symmetry, where $\phi(x)$ turns out 
to be the composite N-G boson having its vev $\langle\phi(x)\rangle_{0}\equiv\phi_{o}=\mu$.  
The free constant $\mu$ introduces the characteristic mass scale $\mu$ which also defines the vev 
of $l_{\mu\nu}$ given by the formula $\langle l_{\mu\nu}(x)\rangle_{0}= (\mu/4)g_{\mu\nu}(x)$. 

At the broken vacuum state marked by $\phi_{o}$ the potential becomes a constant and we identify it with the cosmological constant. As a result, we get the opportunity to control the value of the cosmological constant by varying free dimensionless  parameters of the potential.
Using this possibility we find such two- and four-parametric potentials for which the cosmological constant becomes equal to zero.

The sections of the paper are distributed as follows.
In Sec.1 we present the action, the potential of $l_{\mu\nu}(x)$ in the background with a metric $g_{\mu\nu}(x)$ and define the action symmetries. 
In order not to overload  the paper we restrict ourselves  to the case, where $g_{\mu\nu}$ is considered as an external gravitational field.
In Sec.2 the EOM and its representation including the Riemann tensor $R_{\mu\nu\gamma\rho}$ of the background are derived. 
In Sec.3 spontaneous emergence of the composite scalar field $\phi(x)$ is detected. 
In Sec.4 the non-trival extremal of the two-parametric potential is found out. 
It is shown that on this extremal the action reduces to the scale and diff invariant action of the massless $\phi^{4}$-theory in the background with the coupling constant equal to  $\frac{k^{2}}{2}(\alpha-\beta)$, where $\alpha$ and $\beta$ are free parameters. 
Revealed in Sec.5 is spontaneous breaking of the Weyl and scale symmetries  resulting in the cosmological constant equal to $\lambda= \frac{\alpha-\beta}{32k^{2}}\langle\phi\rangle_{0}^{4}$.  
This value is equal to the potential of the vacuum solution breaking the Weyl and scale symmetries.
In Sec.6 it is shown that the extremal of a four-parametric potential coincides with the found extremal of the two-parametric potential. This result shows that spontaneous breakdown of the Weyl and scale symmetries takes place for the general polynomial potential in four-dimensional gravitational backgrounds. The obtained results and further prospects are discussed in Summary.

\section{Weyl and scale invariant action in a background}

Let us consider the generally covariant action for a massless symmetric tensor field  $l_{\mu\nu}(x^{\rho})$ in a 4-dim. curved background with the metric tensor $g_{\mu\nu}(x)$ 
\begin{eqnarray}\label{actncmc}
S =\frac{1}{k^{2}} \int d^{4}x\sqrt{|g|}
(\frac{1}{2}\nabla_{\mu}l_{\nu\rho}\nabla^{\{\mu}l^{\nu\}\rho}
-\nabla_{\mu}l^{\mu}_{\rho}\nabla_{\nu}l^{\nu\rho} + U(g,l)),
\end{eqnarray}
where the brackets $\{\mu\nu\}, [\mu\nu]$ imply the symmetrization 
and antsymmetrization  $(\mu\nu \pm \nu\mu)$, respectively \cite{Z1}. 
As already noted in the Introduction and will be proved below, the presence of two 
massless tensor fields in (\ref{actncmc}) will not violate the no-go theorem  discussed there.
One can choose 
the scalar potential $U$  as the sum of homogenious quartic polynomials 
invariant under diffeomorphisms with free dimensionless phenomenological constants $\alpha, \beta, b_{2}, b_{4}, b'_{4}$ 
\begin{eqnarray}\label{U}
 U=\frac{2}{3} \,\alpha Trl\,Tr(l^{3}) - \frac{1}{2}\beta(Tr(l^{2}))^{2} + b_{2}Tr(l^{2})\,(Trl)^2 +  b_{4}(Trl)^{4} + b'_{4} Tr(l^{4}).
\end{eqnarray}
The invariants $Tr(l^{n})$ are formed by covariant $g$-contractions of n symmetric tensors $l_{\mu\nu}$ 
\begin{eqnarray}\label{trace}
Trl=l_{\mu\nu}g^{\mu\nu}, \ \  Tr(l^{2})=l_{\mu\nu}l^{\nu\mu}, \ \ \ \ Tr(l^{3})=l_{\mu\rho}l^{\rho\gamma}l_{\gamma}{}^{\mu}, \ \ \ \  Tr(l^{4})=l_{\mu\rho}l^{\rho\gamma}l_{\gamma\lambda}l^{\lambda\mu}.
\end{eqnarray}
This action has a dimensionless coupling $k$ and can be treated as a generally covariant
 extension of  $\phi^{4}$ theory with $\phi \longrightarrow l_{\mu\nu}$. 
The functional  (\ref{actncmc}) is invariant under the global Weyl 
\begin{eqnarray}\label{wtot} 
 x^{\prime}{}^{\mu}= x^{\mu}, \, \, \, \,  \, g_{\mu\nu}^{\prime}(x^{\prime})= 
 e^{2\alpha}g_{\mu\nu}(x),  \, \, \, \,  \, l_{\mu\nu}^{\prime}(x^{\prime})=e^{\alpha}l_{\mu\nu}(x).
\end{eqnarray}
and scaling transformations
\begin{eqnarray}\label{diltot}  \,
 x^{\prime}{}^{\mu}= e^{-\lambda}x^{\mu}, \, \, \, \, \, g_{\mu\nu}^{\prime}(x^{\prime})= 
 g_{\mu\nu}(x),  \, \, \, \,  \, l_{\mu\nu}^{\prime}(x^{\prime})= e^{\lambda}l_{\mu\nu}(x).
\end{eqnarray}

Our goal is to study the shapes of the interaction potential encoding self-interaction of the field $l_{\mu\nu}$ in the gravitational background with the metric $g_{\mu\nu}(x)$. The knowledge of the extremals of $U$ has to give an information about the possibility for spontaneous breaking 
of the above discussed global symmetries.

\section{Equation of motion and its diverse formulations}

The covariant equation of motion  for the dynamical field $l_{\mu\nu}$ following from (\ref{actncmc}) is
\begin{eqnarray}\label{vareq}
\frac{1}{2}\nabla_{\bullet}\nabla^{\{\bullet}l^{\{\nu\}\rho\}}
-\nabla^{\{\nu}  \nabla_{\bullet} l^{\rho\}\bullet}
=
\frac{\partial U}{\partial l_{\nu\rho}} \, ,
\end{eqnarray}
where the notation $\nabla_{\bullet}F^{\bullet \beta...\lambda}$ encodes the Einstein's summing rule: $\nabla_{\bullet}F^{\bullet \beta...\lambda}\equiv\nabla_{\alpha}F^{\alpha \beta...\lambda}$.
Eq. (\ref{vareq}) hides the Riemann tensor $R_{\mu\nu\gamma\rho}$ of the background metric $g_{\mu\nu}$. 
To show the presence of this tensor in (\ref{vareq}) we extract the commutator in its l.h.s. 
\begin{eqnarray}\label{eqrl}
\frac{1}{2}\nabla_{\bullet}\nabla^{[\bullet}l^{\{\nu]\rho\}}
+ [\nabla^{\bullet},  \nabla^{\{\nu}] l_{\bullet}{}^{\rho\}}
= \frac{\partial U}{\partial l_{\nu\rho}}   
\end{eqnarray}
and write (\ref{eqrl}) in the form containing the D'Alembertian $\Box\equiv \nabla_{\mu}\nabla^{\mu}$ in its l.h.s.  
\begin{eqnarray}\label{eqA}
\nabla_{\bullet}\nabla^{\bullet}l^{\nu\rho}
-\frac{1}{2}\nabla^{\{\nu}  \nabla_{\bullet}l^{\rho\}\bullet} 
=
-\frac{1}{2}[\nabla^{\bullet},  \nabla^{\{\nu}] l_{\bullet}{}^{\rho\}}
+ \frac{\partial U}{\partial l_{\nu\rho}}.
\end{eqnarray}
Then the Bianchi identitity 
\begin{eqnarray}\label{BI}
[\nabla_{\mu} , \, \nabla_{\nu}] l^{\gamma\rho}
=R_{\mu\nu}{}^{\gamma}{}_{\lambda} l^{\lambda\rho}  
+ R_{\mu\nu}{}^{\rho}{}_{\lambda} l^{\gamma\lambda}   
\end{eqnarray}
allows to write  the $\nu\rho$-symmetrized commutator in (\ref{eqA}) in the desired form   
\begin{eqnarray}\label{BIcon} 
[\nabla^{\bullet},  \nabla^{\{\nu}] l_{\bullet}{}^{\rho\}}
=
R^{\bullet\{\nu\rho\}\star}{}  \,l_{\star \,\bullet} 
+ R^{\star\{\nu} l^{\rho\}}{}_{\star} \,.
\end{eqnarray}
For compactness,  we write this equation in a condenced form 
\begin{eqnarray}\label{BIcond} 
[\nabla^{\bullet},  \nabla^{\{\nu}] l_{\bullet}{}^{\rho\}}
=
K^{\bullet\{\nu\rho\}\star}{}  \,l_{\star \,\bullet}  
 \end{eqnarray}
introducing the tensor  $K^{\bullet\{\nu\rho\}\star}{}$ which denotes the r.h.s. of (\ref{BIcon}) 
\begin{eqnarray}\label{BIK} 
K^{\bullet\{\nu\rho\}\star}:
=
R^{\bullet\{\nu\rho\}\star}  
+ R^{\bullet\{\nu} g^{\rho\}\star}.
\end{eqnarray}
It is easy to check that $K^{\bullet\{\nu\rho\}\star}{}$ obeys the $g$-traceless condition
\begin{eqnarray}\label{idK}
K^{\alpha\,\bullet}{}_{\bullet}{}^{\beta}:=g_{\nu\rho}K^{\alpha\{\nu\rho\}\beta}=0.                    \end{eqnarray}
As a result, we obtain the representation of EOM (\ref{vareq}) including the D'Alembertian 
 $\Box$ in the curved space-time together with its Riemann and Ricci tensors 
\begin{eqnarray}\label{eqAr}
\nabla_{\bullet}\nabla^{\bullet}l^{\nu\rho}
-\frac{1}{2}\nabla^{\{\nu}  \nabla_{\bullet}l^{\rho\}\bullet} 
=-\frac{1}{2}K^{\bullet\{\nu\rho\}\star}{}  \,l_{\star \,\bullet}
+ \frac{\partial U}{\partial l_{\nu\rho}}.
\end{eqnarray}

Alternatively, one can extract the total covariant divergence in the l.h.s. of EOM (\ref{vareq})
\begin{eqnarray}\label{eqB}
\nabla_{\bullet}\{\nabla^{[\bullet}l^{\nu]\rho}
+ \frac{1}{2}\nabla^{[\nu}l^{\rho]\bullet} \}
=
-[\nabla^{\bullet},  \nabla^{\{\nu}] l_{\bullet}{}^{\rho\}}
+ \frac{\partial U}{\partial l_{\nu\rho}}
\end{eqnarray}
and using (\ref{BIcond}) find another interesting representation for EOM (\ref{vareq}) 
\begin{eqnarray}\label{eqBr}
\nabla_{\bullet}\{\nabla^{[\bullet}l^{\nu]\rho}
+ \frac{1}{2}\nabla^{[\nu}l^{\rho]\bullet} \}
=-K^{\bullet\{\nu\rho\}\star}{}  \,l_{\star \,\bullet} 
+ \frac{\partial U}{\partial l_{\nu\rho}}.
\end{eqnarray}
This representation reveals the presence of  $\nabla_{\mu}l_{\nu\rho}$ in EOM 
in the special combination
\begin{eqnarray}\label{rot}
H^{\mu\nu\rho}:=\nabla^{[\mu}l^{\nu]\rho} 
\end{eqnarray}

Thus, we have the two equivalent formulations (\ref{eqAr}) and (\ref{eqBr}) 
for the EOM  (\ref{vareq}) that will help to investigate the proposed model (\ref{actncmc}).

\section{Spontaneous emergence of composite scalar field}

It is hard to solve  nonlinear Eqs.(\ref{eqAr}) or (\ref{eqBr}). But it is easier to find classical vacua defined by extremals of $U$. To this end we multiply (\ref{eqAr}) by $g_{\nu\rho}$ that  gives 
\begin{eqnarray}\label{eqred} 
\nabla^{\bullet}\{\nabla_{\bullet} Trl - \nabla_{\mu}l^{\mu}{}_{\bullet}\}
=g_{\nu\rho}\frac{\partial U}{\partial l_{\nu\rho}}
\end{eqnarray}
after using (\ref{idK}). It shows that for potentials obeying the condition 
\footnote{ We impose this condition  in order to simplify the analysis of possible solutions of nonlinear EOM  (\ref{eqAr}).  In the next Section we show that (\ref{resr}) leads to  the 
constraints  (\ref{solII}) that reduce the number of independent parameters of the general potential    $U$ (\ref{U}) to two parameters $\alpha $ and $\beta$.}
\begin{eqnarray}\label{resr} 
g_{\nu\rho}\frac{\partial U}{\partial l_{\nu\rho}}=0
\end{eqnarray}
 Eq. (\ref{eqred}) transforms into the continuity equation 
\begin{eqnarray}\label{eqred'} 
\nabla^{\bullet}\{\nabla_{\bullet} Trl - \nabla_{\mu}l^{\mu}{}_{\bullet}\} =0. 
\end{eqnarray}
The latter is presented in the  form of the momentum conservation 
\begin{eqnarray}\label{law}
\frac{1}{\sqrt{|g|}}\partial_{\mu} 
 \{ {\sqrt|g|}
 (g^{\mu\bullet}\partial_{\bullet}\phi 
 - \nabla_{\bullet}l^{\bullet\mu}) \}=0 
\end{eqnarray}
of the spontaneously emerging scalar field $\phi$ accompanied with other components of $l_{\mu\nu}$
 \begin{eqnarray}\label{cond}
\phi:=Trl\equiv g^{\mu\nu}l_{\mu\nu}.
\end{eqnarray}  
Isolating the D'Alambertian $\Box$ from (\ref{law}) gives the generalized Klein-Gordon equation for $\phi$  
\begin{eqnarray}\label{box}
 \Box \phi= \frac{1}{\sqrt{|g|}}\partial_{\mu} 
 ( {\sqrt|g|} \, \nabla_{\bullet}l^{\bullet\mu}),
\end{eqnarray}
where the vector field $\nabla_{\bullet}l^{\bullet\mu}$ might be treated as an effective source. 
If the source is absent, e.g. when  $\nabla_{\bullet}l^{\bullet\mu}=0 $,
 Eq. (\ref{box}) acquires the particular solution  
\begin{eqnarray}\label{vacI}
\nabla_{\bullet}l^{\bullet\mu}=0 \, \, \, \rightarrow \, \, \,  l_{o\nu\rho}=\frac{\mu}{4} g_{\nu\rho} \, \, \, \rightarrow \, \, \, \phi_{o}=\mu, 
\end{eqnarray} 
where $\mu$ is a constant of the dimension $(length)^{-1}$ in the system of units $\hbar=c=1$. The solution (\ref{vacI}) hints that $\phi_{o}$ might be a critical point of $U$ associated with the vev $\langle\phi\rangle_{0}=\mu$. Such a possibility would  signalize about spontaneous breaking of the scale symmetry. This raises the question about extremals of the interaction potential $U$.

\section {Extremals of interaction potential}

To answer this question we study the derivative of $U$    
\begin{eqnarray}\label{derU}
 \frac{1}{2}\frac{\partial U}{\partial l_{\nu\rho}}
 =
 \alpha\,  (l^{2})^{\nu\rho}Trl- \beta \, l^{\nu\rho}\, Tr(l^{2}) + b_{2} \, l^{\nu\rho}\, (Trl)^{2} 
+ 2 b'_{4}\, (l^{3})^{\nu\rho}  \\  
+ \, [\, \frac{\alpha}{3}\, Tr(l^{3}) + \,  b_{2}\, Trl\, Tr(l^{2})
 + \, 2b_{4}\,  (Trl)^{3}\, ]\, \frac{\partial Trl}{\partial l_{\nu\rho}}. \nonumber
\end{eqnarray}
Subsequent substitution of $g$-contracted Eq. (\ref{derU}) in Eq. (\ref{resr}) gives  
 the equation  
\begin{eqnarray}\label{resr'}
 \frac{1}{2}g_{\nu\rho}\frac{\partial U}{\partial l_{\nu\rho}}
 =
 (\alpha -\beta) TrlTr(l^{2}) + b_{2} \,Trl(Trl)^{2} + 2 b'_{4}\, Tr(l^{3})
  \\  
+ \, [\, \frac{\alpha}{3}\, Tr(l^{3}) + \,  b_{2}\, Trl\, Tr(l^{2})
 + \, 2b_{4}\,  (Trl)^{3}\, ]\, (g_{\nu\rho}\frac{\partial Trl}{\partial l_{\nu\rho}}) =0. \nonumber
\end{eqnarray}
Then the substitution of  $\frac{\partial Trl}{\partial l_{\nu\rho}}=g^{\nu\rho}$
in (\ref{resr'}) yields the algebraic equation 
\begin{eqnarray}\label{II}
 (\frac{4}{3}\alpha +2 b'_{4})\, Tr(l^{3}) + [(\alpha -\beta)+ 4b_{2}]TrlTr(l^{2})
+ (b_{2} +8b_{4})(Trl)^{3}=0
\end{eqnarray}
which is solved by the choice of the three phenomenological constants 
\begin{eqnarray}\label{solII}
b'_{4}=-\frac{2}{3}\alpha, \, \, \,  b_{2}=\frac{1}{4}(\beta-\alpha), \, \, \,
b_{4}=- \frac{1}{8} b_{2}.
\end{eqnarray}
Thus, we obtain the two-parametric homogenious potential
\begin{eqnarray}\label{UII}
U_{II}=\frac{2}{3}\alpha [ TrlTr(l^{3})-Tr(l^{4})] - \frac{1}{2}\beta (Tr(l^{2}))^{2} 
+ \frac{1}{4}(\beta-\alpha) [Tr(l^{2})(Trl)^{2} - \frac{1}{8}(Trl)^{4}]
\end{eqnarray}
and its partial derivative with respect to $l_{\nu\rho}$
\begin{eqnarray}\label{derUII}
\frac{\partial U_{II}}{\partial l_{\nu\rho}}
 =
\frac{2}{3}\alpha [3(l^{2})^{\nu\rho}Trl +  g^{\nu\rho}Tr(l^{3}) - 4(l^{3})^{\nu\rho}]   
-2\beta l^{\nu\rho}Tr(l^{2})
 \\
+ \frac{1}{2}(\beta-\alpha) [l^{\nu\rho}(Trl)^{2} 
+ g^{\nu\rho}TrlTr(l^{2}) -\frac{1}{2}g^{\nu\rho}(Trl)^{3}]. \nonumber
\end{eqnarray}
The latter expression is  the cubic polynomial in $l_{\nu\rho}$ 
\begin{eqnarray}\label{derUIIm}
\frac{\partial U_{II}}{\partial l_{\nu\rho}} 
 =
 \{- \frac{8}{3}\alpha(l^{3})^{\nu\rho} + 2\alpha Trl (l^{2})^{\nu\rho}
 + \, [\frac{1}{2}(\beta-\alpha)(Trl)^{2} - 2\beta Tr(l^{2})] \, l^{\nu\rho} \}   
  \\ 
  +   \{ \frac{2}{3}\alpha Tr(l^{3}) + \frac{1}{2}(\beta-\alpha)[Tr(l^{2})
- \frac{1}{4}(Trl)^{2}]\, Trl \} \,  g^{\nu\rho}
 \nonumber
\end{eqnarray}
spanned by the tensors $g^{\nu\rho}, \, l^{\nu\rho}$ and their products.
It is easy to check that (\ref{derUIIm}) obeys the constraint (\ref{resr}) 
and the homogenuity condition
\begin{eqnarray}\label{verif}
g_{\nu\rho}\frac{\partial U_{II}}{\partial l_{\nu\rho}}=0, \, \, \, \, \, \,
l_{\nu\rho}\frac{\partial U_{II}}{\partial l_{\nu\rho}}=4U_{II}. 
\end{eqnarray}
Then the extremals of the  potential (\ref{UII}) are given  by zeroes of Eq. (\ref{derUIIm})
\begin{eqnarray}\label{ZerUIIm}
\{- \frac{8}{3}\alpha(l^{3})^{\nu\rho} + 2\alpha \phi (l^{2})^{\nu\rho} 
 + \, [\frac{1}{2}(\beta-\alpha)\phi^{2} - 2\beta Tr(l^{2})] \, l^{\nu\rho} \}   
  \\ 
  +   \{ \frac{2}{3}\alpha Tr(l^{3}) + \frac{1}{2}(\beta-\alpha)[Tr(l^{2})
- \frac{1}{4}\phi^{2}]\, \phi \} \,  g^{\nu\rho} =0, \nonumber
\end{eqnarray}
where $\phi=Trl$ (\ref{cond}). 
For the case $\alpha=0$ Eq. (\ref{ZerUIIm}) factorises in the product 
\begin{eqnarray}\label{ZerU1}
\frac{\partial U_{II}}{\partial l_{\nu\rho}}|_{\alpha=0} 
 =-2\beta l_{\alpha\beta}(l^{\alpha\beta} -\frac{\phi}{4}g^{\alpha\beta}) 
 (l^{\nu\rho} -\frac{\phi}{4}g^{\nu\rho} )=0
\end{eqnarray}
 which shows that we have three roots. The first of them is the zero root    
\begin{eqnarray}\label{alphbet}
l_{1 \mu\nu}=\phi_{1}=0. 
\end{eqnarray} 
but the second double degenerate root is 
\begin{eqnarray}\label{alph0}
 l_{0\mu\nu}(x)= \frac{Trl_{0}(x)}{4}g_{\mu\nu} \equiv \frac{\phi_{0}(x)}{4}g_{\mu\nu}.
 \end{eqnarray}
So, we obtain  the reduced one-parametric potential 
\begin{eqnarray}\label{UIIa0}
 U_{II}|_{\alpha=0}
 = - \frac{1}{2}\beta \{(Tr(l^{2}))^{2} - \frac{1}{2} [Tr(l^{2})(Trl)^{2} - \frac{1}{8}(Trl)^{4}]\}
\end{eqnarray}
with the one non-trivial extremal (\ref{alph0}).  The exclusive property of this matrix is that its nth power is equal to itself multiplied by $(\frac{\phi_{0}}{4})^{n-1}$
\begin{eqnarray}\label{degr}
(l_{0}^{n})_{\mu\nu}=(\frac{\phi_{0}}{4})^{n} g_{\mu\nu}\equiv (\frac{\phi_{0}}{4})^{n-1}l_{0\mu\nu},
 \, \, \, \, \, \, \,
Tr(l_{0}^{n})=4(\frac{\phi_{0}}{4})^{n}\equiv(\frac{\phi_{0}}{4})^{n-1}Trl_{0}.   
\end{eqnarray} 
 Due to this, $l_{0\mu\nu}$ turns out to be a root of Eq. (\ref{ZerUIIm}) with $\beta=0$ as seen   from the representation 
\begin{eqnarray}\label{UIa0}
 \frac{\partial U_{II}}{\partial l_{\nu\rho}}|_{\beta=0}=
-\alpha\{ \frac{8}{3}\,[(l^{3})^{\nu\rho} - \frac{1}{4} Tr(l^{3})g^{\nu\rho}]  
\\ 
-2Trl\, [(l^{2})^{\nu\rho} - \frac{1}{4}Trl \,l^{\nu\rho}]  + \frac{1}{2}Trl \,[Tr(l^{2}) -\frac{1}{4}(Trl)^{2}]\, g^{\nu\rho} \}=0, \nonumber
\end{eqnarray}
 obtained by regrouping of the terms in (\ref{ZerUIIm}). Indeed, each term in square brackets in (\ref{UIa0}) vanishes after substitution of $l_{0\mu\nu}$ there. 
 So, we obtain  another one-parametric potential
\begin{eqnarray}\label{UIIb0}
U_{II}|_{\beta=0}
= -\alpha\{\frac{2}{3}[Tr(l^{4})- Trl Tr(l^{3})] 
+ \frac{1}{4} [Tr(l^{2})(Trl)^{2} - \frac{1}{8}(Trl)^{4}]\}.
\end{eqnarray}
Thus, we see that the two-parametric potential (\ref{UII}) 
\begin{eqnarray}\label{UIIs}
U_{II}= \beta U_{II}|_{\alpha=0} + \alpha U_{II}|_{\beta=0}
\end{eqnarray}
has the same extremals as $U_{II}|_{\alpha=0}$ and $U_{II}|_{\beta=0}$.
 At the extremal (\ref{alph0}) $U_{II}$ takes the form
\begin{eqnarray}\label{UIIexval}
U_{II}|_{l=l_{0}}=\frac{1}{32}(\alpha-\beta) \phi_{0}^{4} 
\end{eqnarray}
due to the relations between the polynomial values at $l_{0\mu\nu}$
\begin{eqnarray}\label{Uvac}
\frac{2}{3} Trl_{0}\,Tp(l_{0}^{3})-\frac{1}{24}(Trl_{0})^4 
=\frac{1}{2}(Tr(l_{0}^{2}))^{2}- \frac{1}{32}(Trl_{0})^4=0,  \\
Tr(l_{0}^{2})(Trl_{0})^2-\frac{1}{8}(Trl_{0})^4=\frac{1}{8}(Trl_{0})^4,\, \, \,
Tp(l_{0}^{4})- \frac{1}{64}(Trl_{0})^4=0. \nonumber
\end{eqnarray} 
When $\alpha=\beta$ the potential (\ref{UIIs}) transforms to the polynomial  
\begin{eqnarray}\label{UIIab}
U_{II}|_{\alpha=\beta} 
=\alpha \{\frac{2}{3} TrlTr(l^{3})- \frac{1}{2} (Tr(l^{2}))^{2} - \frac{2}{3} Tr(l^{4}) \} 
\end{eqnarray}
and vanishes on the extremal $l_{o\mu\nu}$
\begin{eqnarray}\label{UIIo ab}
U_{II}|_{\alpha=\beta; \, l=l_{o}} =0.
\end{eqnarray}
Going back to the expression for the action  (\ref{actncmc}) at $l_{0\mu\nu}(x)$ 
we find its kinetic term 
\begin{eqnarray}\label{kin4}
(\frac{1}{2}\nabla_{\mu}l_{\nu\rho}\nabla^{\{\mu}l^{\nu\}\rho}
-\nabla_{\mu}l^{\mu}_{\rho}\nabla_{\nu}l^{\nu\rho})|_{l=l_{0}}=\frac{1}{4}\partial_{\mu}\phi_{0}\partial^{\mu}\phi_{0}.
\end{eqnarray}
Then $S$ (\ref{actncmc}) transforms in the scale  and diff-invariant action of the massless $\phi^{4}$ theory 
\begin{eqnarray}\label{actnUII0}
S_{II0} = \int d^{4}x\sqrt{|g|}(\partial_{\mu}\tilde{\phi_{0}}\partial^{\mu}\tilde{\phi_{0}} +  \frac{k^{2}}{2}(\alpha-\beta) \tilde{\phi}_{0}^{4}),
\end{eqnarray}
in the background, where we substitute $U_{II}|_{l=l_{0}}$ (\ref{UIIexval}) and redifine $\phi_{0}$ in $\tilde{\phi}_{0}:=(2k)^{-1}\phi_{0}$.
 
 Thus, we build the potential function  $U_{II}$ with the extremals (\ref{alphbet}) and  (\ref{alph0}). The former extremal is a particular solution of EOM (\ref{eqAr}) and (\ref{eqBr}), but this property should be verified for  $l_{0\, \mu\nu}(x)$ to confirm the assumption made in Section 3.
 Let us note that the potential term in (\ref{actnUII0}) vanishes for the case when the free parameters $\alpha$ and $\beta$ are equal. Below this fact will be used in the proof of the vanishing of the cosmological constant under spontaneous breaking of the scale symmetry.

\section{Breaking of the Weyl and scale symmetries}

To check whether the extremal (\ref{alph0}) is a solution of EOM (\ref{eqAr}) we note that 
the expression under the covariant derivative in Eq. (\ref{eqred'}) reduces to the derivative 
of $ \phi_{0}$  
\begin{eqnarray}\label{vacrel}
\nabla_{\rho} Trl_{0} - \nabla_{\bullet}l^{\bullet}_{0\rho}
=\frac{3}{4}\partial_{\rho}\phi_{0}
\end{eqnarray}
due to the relation
\begin{eqnarray}\label{totder}
\nabla_{\bullet}l_{0}^{\bullet\mu}= \frac{1}{4}g^{\mu\bullet}\partial_{\bullet}\phi_{0}.
\end{eqnarray} 
Then Eq. (\ref{eqred'}) itself is transformed into the D'Alembert equation in the curve space
\begin{eqnarray}\label{Dalm}
\nabla^{\rho} \{\nabla_{\rho} Trl_{0} - \nabla_{\bullet}l^{\bullet}_{0\rho}\}
=\frac{3}{4}\Box \phi_{0}(x)=0  \, \, \, \longrightarrow \, \, \, \Box \, l_{0\mu\nu}(x)=0, 
\end{eqnarray}
since the second term in the l.h.s. of (\ref{Dalm}) becomes proportional to $\Box \phi_{0}(x)$    
\begin{eqnarray}\label{N-Geq} 
\nabla^{\rho} \nabla_{\bullet}l^{\bullet}_{0\rho}=\frac{1}{4}\Box \phi_{0}(x).
\end{eqnarray}
The operator $\Box$ is obtained from $\partial_{\bullet}\partial^{\bullet}$ by the replacement of the flat metric $\eta_{\mu\nu}$ in $g_{\mu\nu}(x)$ 
\begin{eqnarray}\label{boxNG}
\Box \phi_{0}:= \nabla^{\bullet}\nabla_{\bullet}\phi_{0}=g^{\mu\nu}\partial_{\mu}\partial_{\nu}\phi_{0} -  g^{\mu\nu}\Gamma^{\rho}_{\mu\nu}\partial_{\rho}\phi_{0}. \, \, \, \, \, \, \, \, \, \, \, \, 
\end{eqnarray}  
On the extremals of $U_{II}$ EOM (\ref{eqAr}) reduces to   
\begin{eqnarray}\label{eqAr0}
\nabla^{\{\nu}  \nabla_{\bullet}l_{0}^{\rho\}\bullet} 
=
K^{\bullet\{\nu\rho\}\star}{}  \,l_{0\,\star \,\bullet}
\end{eqnarray}
due to  Eq. (\ref{Dalm}). 
Inserting (\ref{totder}) in (\ref{eqAr0}) and using  (\ref{BIcond}), that defines   $K^{\bullet\{\nu\rho\}\star}{}  \,l_{0\star \,\bullet}$,  gives  
\begin{eqnarray}\label{eqAr1}
g^{\bullet \{\rho} \nabla^{\nu \}}\partial_{\bullet}\phi_{0}= 
K^{\bullet\{\nu\rho\}}{}_{\bullet}\phi_{0}
=(R^{\bullet\{\nu\rho\}}{}_{\bullet} + R^{\{\nu\rho\}})\phi_{0}.
\end{eqnarray}
Then taking into account the identity 
\begin{eqnarray}\label{Rim0}
R^{\bullet\{\nu\rho\}}{}_{\bullet}=-R^{\{\nu\rho\}} \, \, \,\longrightarrow \, \, \,
(R^{\bullet\{\nu\rho\}}{}_{\bullet} + R^{\{\nu\rho\}})\equiv 0
\end{eqnarray}
  the r.h.s. of Eq. (\ref{eqAr1}) vanishes, because 
 \begin{eqnarray}\label{eqAr00}
 K^{\bullet\{\nu\rho\}\star}{}  \,l_{0\,\star \,\bullet}=0.
 \end{eqnarray}\
So, EOM  (\ref{eqAr1}) on the extremal (\ref{alph0}) is simplified to the equation 
 \begin{eqnarray}\label{eqAr1r}
 g^{\bullet \{\rho} \nabla^{\nu \}}\partial_{\bullet}\phi_{0}=0. 
 \end{eqnarray}
and its  multiplication by $g_{\nu\rho}$ brings us back to the Klein-Gordon equation (\ref{Dalm})  
\begin{eqnarray}\label{eqAr2}
 g_{\nu\rho}g^{\bullet \{\rho} \nabla^{\nu \}}\partial_{\bullet}\phi_{0}=2\Box \phi_{0}= 0. 
 \end{eqnarray}
 In the explicit form  Eq. (\ref{eqAr1r}) is written as 
 \begin{eqnarray}\label{eqA3} 
 g^{\rho \{\bullet} g^{\star\}\nu }[\partial_{\star}\partial_{\bullet}\phi_{0} 
 - \Gamma^{\alpha}_{\bullet\star}\partial_{\alpha}\phi_{0}]= 0 \, \, \,\longrightarrow \, \, \,
 \partial_{\alpha}\partial_{\beta}\phi_{0} 
 - \Gamma^{\bullet}_{\alpha\beta}\partial_{\bullet}\phi_{0}=0.
 \end{eqnarray}
 The simplest covariant  solution of both Eq. (\ref{eqAr1r}) and the Klein-Gordon Eq. (\ref{eqAr2}) is 
 \begin{eqnarray}\label{muvac}
  \phi_{0}(x)  \, \, \, \longrightarrow \, \, \,   \phi_{o}=\mu=constant.
 \end{eqnarray}
This solution confirms the conjecture of Section 3 on the solution (\ref{vacI}) to be an extremal of $U$ which in its  turn proves spontaneous breakdown of both the Weyl and scale symmetries. Indeed, the extremal (\ref{alph0}) of $U_{II}$, realized by the propagating field $l_{0\mu\nu}(x)$, transforms  as
\begin{eqnarray}\label{wvac} 
 \phi_{0}^{\prime}(x^{\prime})=e^{-\alpha}\phi_{0}(x) \, \, \longrightarrow \, \,  
 l_{0\mu\nu}^{\prime}(x^{\prime})=e^{\alpha}l_{0\mu\nu}(x)
\end{eqnarray}
under the Weyl tranformations (\ref{wtot}) and 
\begin{eqnarray}\label{dilvac} 
 \phi_{0}^{\prime}(x^{\prime})=e^{\lambda}\phi_{0}(x) \, \, \longrightarrow \, \,  
 l_{0\mu\nu}^{\prime}(x^{\prime})=e^{\lambda}l_{0\mu\nu}(x)
\end{eqnarray}    
under the scale transformations (\ref{diltot}). 
The laws (\ref{wvac}-\ref{dilvac}) 
coincide with those of $l_{\mu\nu}(x)$ itself and hence realize the action symmetries 
on the extremal $l_{0\mu\nu}(x)$ which is  their linear representations. 
Therefore, the orbit swept by $l_{0\mu\nu}(x)$ under the action of the Weyl and scale 
transformations gives rise    to  an infinitely degenerate vacuum manifold. 
 The requirement for  $l_{0\mu\nu}(x)$ to obey EOM (\ref{eqAr1r}) removes this degeneracy 
 by choosing the vacuum solution (\ref{muvac}) 
\begin{eqnarray}\label{vacmu}
\phi_{o}=\mu  \, \, \, \longrightarrow \, \, \,   l_{o\nu\rho}(x)=\frac{\mu}{4} g_{\nu\rho}(x) 
\end{eqnarray}   
 covariant under diffeomorphisms, but breaking the symmetries (\ref{wvac}-\ref{dilvac}).
 Thus, one can treat $\phi(x)$ as the Nambu-Goldstone boson generated by the simultaneous breakdown of the Weyl and scale symmetries 
and having the non-zero vacuum expectation value $\langle\phi\rangle_{0}$  
\begin{eqnarray}\label{scale}
\langle\phi\rangle_{0}\equiv Trl_{o}=\mu.
\end{eqnarray}
It brings the mass scale $\mu$ in the considered model (\ref{actncmc}) with the potential term  $U=U_{II}$.

On the solution (\ref{vacmu}) the kinetic term (\ref{kin4}) vanishes 
and $U_{II}$ is converted to the constant 
\begin{eqnarray}\label{UIIcosm}
U_{II}|_{l=l_{o}}=\frac{1}{32}(\alpha-\beta) \mu^{4}.
\end{eqnarray}
The resulting vacuum action takes the form 
\begin{eqnarray}\label{actnUII}
S_{IIo}=\frac{\mu^{4}}{32k^{2}}(\alpha-\beta) \int d^{4}x\sqrt{|g|}
=\frac{1}{k^{2}} \int d^{4}x\sqrt{|g|}U_{II}|_{l=l_{o}}
\end{eqnarray}
corresponding to the contribution of the cosmological constant
\begin{eqnarray}\label{cosmc}
\lambda_{II}= \frac{\alpha-\beta}{32k^{2}}\langle\phi\rangle_{0}^{4}.
\end{eqnarray}
This expression shows  that on the classical level the cosmological constant 
is proportional to the potential of the classical vacuum state and vanishes when  
$\alpha=\beta$.

\section{Four-parametric potential with broken symmetry}

The results of Sections 4 show that the potential $U$ (\ref{U}) has the  same extremals (\ref{alphbet}), (\ref{alph0}) as the potential (\ref{UIIs}) through the relations (\ref{degr}).
 The latter  substituted in Eq. (\ref{derU}) yield the condition for the phenomenological constants necessary for $l_{0\nu\rho}(x)$ to be an extremal of $U$
\begin{eqnarray}\label{derU0}
 \frac{1}{2}\frac{\partial U}{\partial l_{\nu\rho}}|_{l=l_{0}}
 =(\frac{1}{12}\alpha -  \frac{1}{16}\beta + \frac{1}{2}b_{2} + 2b_{4} + \frac{1}{32}b'_{4})\, \phi_{0}^{3}(x)g^{\nu\rho}(x)=0.
\end{eqnarray}
Eq. (\ref{derU0})is satisfied if $\phi_{0}=0$ or the constants obey the relation
\begin{eqnarray}\label{condco}
\frac{1}{12}\alpha -  \frac{1}{16}\beta + \frac{1}{32}b'_{4} + \frac{1}{2}b_{2} + 2b_{4} =0
\end{eqnarray}
which leaves freedom in the choice of four parameters.
 One can solve (\ref{condco}) expressing, for example, $b_{4}$ in terms of the remaining constants 
\begin{eqnarray}\label{b4}
 b_{4}=-\frac{1}{24}\alpha +  \frac{1}{32}\beta -\frac{1}{64} b'_{4} - \frac{1}{4}b_{2}
\end{eqnarray}
that results in the following four-parametric potential 
\begin{eqnarray}\label{UI}
 U_{I}=\frac{2}{3} \,\alpha Trl\,Tr(l^{3}) - \frac{1}{2}\beta(Tr(l^{2}))^{2} + b_{2}Tr(l^{2})\,(Trl)^2 \\
+  b'_{4} Tr(l^{4}) -  [\frac{1}{24}\alpha - \frac{1}{32}\beta + \frac{1}{64}b'_{4} +  \frac{1}{4}b_{2}](Trl)^{4}. \nonumber
\end{eqnarray}
This expression can be rewritten in a more suitable form
\begin{eqnarray}\label{UI'}
 U_{I}=\alpha[\frac{2}{3}Trl\,Tr(l^{3})- \frac{1}{24}(Trl)^{4}] -  \beta[\frac{1}{2}(Tr(l^{2}))^{2}- \frac{1}{32}(Trl)^{4}] \\
+  b'_{4}[Tr(l^{4})-\frac{1}{64}(Trl)^{4}] +  b_{2}[Tr(l^{2})\,(Trl)^2 - \frac{1}{4}(Trl)^{4}]  \nonumber
\end{eqnarray}
which gives the zero value of the cosmological constant $\lambda$
\begin{eqnarray}\label{UIvac}
U_{I}|_{l=l_{o}}=0 \, \,  \, \, \longrightarrow \, \,  \lambda_{I}=0
\end{eqnarray}
for any choice of $\alpha, \beta, b_{2}$ and $ b'_{4}$ in contrast 
 to (\ref{cosmc}) given by the potential $U_{II}|_{l=l_{o}}$(\ref{UIIcosm}) which  
 was derived using the constraint (\ref{resr}).

So for example choosing
$ b_{2}=b'_{4}=0$ in (\ref{UI}) we get the two-parametric potential
\begin{eqnarray}\label{U1'}
U_{I'}=\frac{2}{3}\alpha Trl\,Tr(l^{3}) -  \frac{1}{2}\beta (Tr(l^{2}))^{2} -[\frac{\alpha}{24}-  \frac{\beta}{32}] (Trl)^{4}
\end{eqnarray} 
which unlike the potential (\ref{UII}) does not contain the monomials $Tr(l^{4})$ and $Tr(l^{2})(Trl)^{2}$.
It is explained by using (\ref{resr}) under the derivation of  $U_{II}$. If  $\alpha=\beta$ then (\ref{U1'}) reduces to
\begin{eqnarray}\label{U1'0}
U_{I'}|_{\alpha=\beta}=\alpha \{\frac{2}{3} Trl\,Tr(l^{3}) - \frac{1}{2}(Tr(l^{2}))^{2}  
- \frac{1}{96}(Trl)^{4}\} 
\end{eqnarray} 
and vanishes on the vacuum extremal (\ref{vacmu}) together with the cosmological constant $\lambda_{I'}$
\begin{eqnarray}\label{UI''0}
U_{I'}|_{\alpha=\beta; \, l=l_{o}}= 0   
 \, \,  \, \, \longrightarrow \, \,  
\lambda_{I'}= 0
\end{eqnarray} 
like $U_{II}|_{\alpha=\beta; \, l=l_{o}}$ in (\ref {UIIo ab}) and $\lambda_{II}$ in (\ref{cosmc}).

So, we have constructed the four-parametric potential (\ref{UI}),  which gives the zero cosmological constant on its extremal $l_{0\mu\nu}(x)$ (\ref{alph0}) and on the vaccuum extremal $l_{o\mu\nu}(x)$, respectively. 
As a result, we have two- and four-parametric potentials for which the cosmological constant becomes equal to zero. These potentials, together with the above built one-parameter potentials, make it possible to control the value of the cosmological constant.

\section{Summary}

 Studied is an attempt to understand the amazing efficiency of scalar fields such as the inflaton in  describing the expanding universe. Our approach is based on the assumption about a composite structure of such scalar particles. The latters are treated as Nambu-Goldstone bosons, which arise as a result of spontaneous breaking of global internal or space-time symmetries.
As an example, a model of the massless symmetric tensor field $l_{\mu\nu}(x)$ in a 4-dim. background $g_{\mu\nu}(x)$ with spontaneously broken global Weyl and scale symmetries is considered. 
It is assumed that such type tensor fields could describe some degrees of freedom of 
the quark-gluon plasma associated with higher spins. 

To explain the proposed scenario, we analyze the critical points of the potential of $l_{\mu\nu}$, represented by general quartic polynomial, and find a classical vacuum that points to spontaneous breaking of both the Weyl and scale symmetries. The vacuum-related  extremal is characterized by the emerging composite Nambu-Goldstone boson $\phi(x):= g^{\mu\nu}l_{\mu\nu}$ which acquires a non-zero vev $\langle\phi\rangle_{0}=\mu$. 
The arbitrary constant $\mu$ introduces a characteristic mass scale which defines the vev 
$\langle l_{\mu\nu}(x)\rangle_{0}= (\mu/4)g_{\mu\nu}(x)$. We show that $\mu$, together with the free  dimensionless parameters of the potential, defines the cosmological constant. 
As a result, it becomes possible to control the value of the cosmological constant by varying 
these free parameters. We find, in particular, that the cosmological constant vanishes for a certain choice of the free parameters. 
In our example the composite structure of the tensor $l_{\mu\nu}$ 
confirms the known no-go theorem which forbids the presence of multiple massless particles with spins equal to 2 in any general covariant action consistent with the perturbative unitarity condition. 
Nevertheless, the theorem allows the field $l_{\mu\nu}$ to carry spin(s) 1 and 0.
 However, spontaneous breaking of the global symmetries, preserving the diff-invariance of the action, demands the extremal $l_{0\mu\nu}(x)$ to be aligned along the external field $g_{\mu\nu}(x)$ similarly to the case of (anti)ferromagnetics in magnetic field. Therefore the polarization DOF of $l_{0\mu\nu}(x)$ are determined by the metric background tensor  $g_{\mu\nu}(x)$. 
 As a result, the $g$-trace of the extremal $l_{0\mu\nu}$ remains its only degree of feedom identified with the composite scalar field. 
 This fact points to the possible distinguished role of composite massless scalar fields in the dynamics of the expanding universe.
 
 It was noted in the Introduction that $l_{\mu\nu}(x)$ is treated as belonging to tensor modes 
generated by the quark-gluon plasma. The scale invariant potential (\ref{U}) captures the dynamics of this mode in the approximation of massless quarks that results in creation of the composite N-G boson $\phi$. In QCD and Standad Model a finite  radius of action of forces binding quarks and gluons is taken into account by introducing the energy scale parameter $\Lambda_{QCD}$. 
The experimentally observed quark masses are expressed in terms of $\Lambda_{QCD}$. 
The pion mass, in particular, is defined by the relation  $m_{\pi}\propto \sqrt{2m_{u}\Lambda_{QCD}}$, 
where $m_{u}$ is the mass of the light current u-quark. The relatively small mass of pion
 in comparison with other mesons is explained by the obsevation that pion is N-G boson of spontaneously broken chiral symmetry. 

Taking into account that in our universe conformal symmetry is broken, one can try to implement a similar mechanism for understanding the emergence of mass for the discussed N-G fields. To do this, note  that the expression  $\Lambda_{QCD}= \langle u\bar{u}\rangle_{0}/f^{2}_{\pi}|_{m_{u}=0}$ contains two dimensional parameters which are the quark condensate and the pion decay constant $f_{\pi}$.  The role of the quark condensate in our model plays  $\langle\phi\rangle_{0}\equiv Trl_{o}=\mu$. Therefore, we need to have a new dimensional constant $f$ encoding an effective mass of $l_{\mu\nu}$. 
Using $f$ allows to extend the scale invariant potential $U$ (\ref{U}) by the additional
  term quadratic in $l_{\mu\nu}$
\begin{eqnarray}\label{Ubr}
U_{broken}=af^{2}Tr(l^{2}) + \frac{b}{4}f^{2}(Trl)^{2} 
\end{eqnarray}
which explicitly breaks the scale symmetry and contains two dimensionless parameters. 
On the extremal (\ref{alph0}) $U_{broken}$ takes the form of the mass term $U_{m}$
for the N-G boson $\phi_{0}(x)$
\begin{eqnarray}\label{Um0}
  U_{broken} \ \ \   \longrightarrow    \ \ \     U_{m}=\frac{(a + b)f^{2}}{4}\phi_{0}^{2}.
\end{eqnarray}
 The sum of  $U$ (\ref{U}) and $U_{broken}$ (\ref{Ubr}) yields an effective potential $U^{*}$
\begin{eqnarray}\label{U*}
U^{*}= U + U_{broken}
\end{eqnarray}
 that explicitly violates the Weyl/scale symmetries but retains the memory of the N-G mechanism of the appearance of the composite scalar $\phi(x)$. 

Another important generalization of the model is the construction of the kinetic term for the  gravitational field. In the theory of gravity the kinetic term for $g_{\mu\nu}$ is given by the Hilbert-Einstein action including the scalar curvature $R$. In the studied model, the requirement
 that the action be invariant under diffeorphisms is extended to its invariance under scaling transformations. These requirements are satisfied by the terms 
linear in $R_{\mu\nu\gamma\rho}$ and quadratic in $l_{\alpha\beta}$. 
Such terms naturally generalize the term $R\phi^2$ used in the previous gravitational models involving the elementary scalar $\phi(x)$. 
These extensions of the model will be studied in more detail in a different place.

 \section{Asknowledgments}

The author is grateful to NORDITA and Physics Department of Stockholm University for kind hospitality and support. I also thank D. Uvarov, H. von Zur-M\"{u}hlen, F. Wilczek for helpful discussions, as well as  A. Davidson and A. Krikun for useful communications.

\end{document}